# Convex hull: Incremental variations on the Akl-Toussaint heuristics
## Simple, optimal and space-saving convex hull algorithms

**Jean Souviron[1]**

COGITECH Jean Souviron
613 D'Ailleboust
Montréal, Quebec,
H2R 1K2 Canada

*JeanSouviron@hotmail.com*

**Abstract :** Convex hulls are a fundamental geometric tool used in a number of algorithms. A famous paper by Akl & Toussaint in 1978 described a way to reduce the number of points involved in the computation, which is since known as the *Akl-Toussaint heuristics*. This paper first studies what this heurstics really represents in terms of reduction of points and demonstrates that the optimum selection is reached using an octogon as the remaining number of points is in $O(\sqrt{N})$ rather than the usual $O(N)$. Then it focuses on optimising the overall computational *efficiency* in a convex hull computation. Although the heuristics is usually used as a first step in computations one can obtain the convex hull directly from the heuristics's basis. First a simple incremental implementation is described, and if the number of characteristic points of the Akl-Toussaint heuristics $p$ is taken as a parametre the convex hull is then computed in a $O(N(p+h/p))$ average complexity or $O(Nh)$ asymptotic complexity. Given the relative constant factor of $1/p$ however experimental results show that this algorithm should be considered linear in average. Worst-case complexity is in $O(N^2)$ and space complexity is $O(h)$ but could be $O(1)$ if the required output is the array of convex vertices's indexes. Then a remark on why the basic incremental method should be preferred for average cases is made. Finally an optimal linear algorithm both in average and worst-case and using a minimal space complexity in $O(\sqrt{N})$ in average (*or O(1) if in-place computation is allowed*) is presented.

---

[1] Dr Jean Souviron, Ph.D 1984, is an independent consultant in scientific programming since 1994..

# 1. Introduction

Convex hulls are the stepping stone of a number of algorithms, mainly because of their fast geometrical delimitation of the space occupied by the points. As such they are one of the most researched subject in computational geometry as well as a prolific publication field. No attempt to build an exhaustive bibliography on convex hull computation will be made as the sheer number of publications, algorithms, and lists already existing provides the reader with a vast pool of references.

Several well-documented surveys are available on the Web such as "*A History of Linear-time Convex Hull Algorithms for Simple Polygons* " by Aloupis[2], "*Computational Geometry on the Web*" by Toussaint[15], or "*A Case Study on the Cost of Geometry Computing*" by Schirra[14]. Linear complexity can be achieved in computing the convex hull of simple polygons, as a variety of references in Aloupis show. However for a general set of unordered points no linear algorithm exists. The most well-known algorithms for general sets of points are the Incremental approach (*Kallay[12]*), the Jarvis'March (*Jarvis[11]*), the Graham Scan (*Graham[10]*), the QuickHull (*independently by Eddy[7] and Bykat[4]*), the Kirkpatrick-Seidel algorithm (*Kirkpatrick & Seidel[13]*) and finally Chan's algorithm (*Chan[5]*).

In 1978 a simple yet efficient algorithm was published and part of it is now known as *the Akl-Toussaint heuristics*[1]. Its goal is to drastically reduce the number of points involved in the computations by defining a first approximate of the convex hull based on some characteristic points known to be on the hull and easy to compute. However this heuristics provides only for a way of *eliminating* points, letting the choice of the algorithm to compute the remaining of the convex hull open. For instance the authors use a Divide-And-Conquer strategy to compute separate convex hulls on the subsets, using for instance a Graham Scan, and then merge them (*this strategy is also mentioned for instance in Chan's paper*).

This heuristics is widely used as a means to reduce the number of points before a convex hull computation. This paper first studies the heuristics and then presents an algorithm using its mechanism to *directly* compute the convex hull.

## 2. The Akl-Toussaint heuristics principle

The implementation of the Akl-Toussaint heuristics is in two steps:

- Compute the pre-defined limiting points, thus defining a box
- Remove the points which are inside this box

### 2.1 Defining the initial points at the root of the Akl-Toussaint heuristics

In their original paper Akl & Toussaint used a quadrilateral formed by the extremes of each coordinate. This is also what is usually used in the industry and research worlds (*e.g. CGAL*). However later suggestions arose that the number of initial points could be optionally increased to include either or both of the points with smallest and largest sums of x- and y-coordinates or those with smallest and largest differences of x- and y-coordinates, as they all also are for sure on the convex hull. These three different cases were investigated in terms of their efficiency at removing points. In order to run these tests computer-generated random data were used.

The random-generated data are of three kinds: rectangle-bound and either uniformly or centrally distributed circle-bound data. Examples of each are shown below.

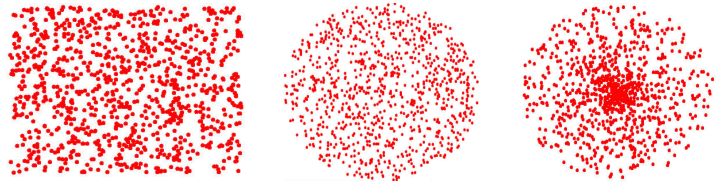

**Figure 1.** The three types of random data used to test progams

Figure 2 shows a comparison of results for the three kinds of random data first using the usual quadrilateral, then an hexagon based on the sum of coordinates and finally an octogon.

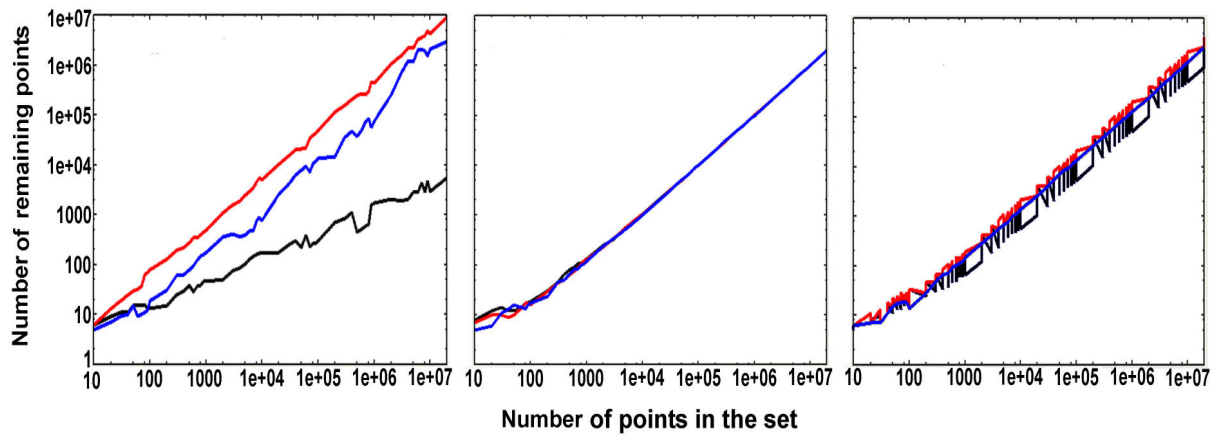

**Figure 2.** Number of points left after Akl-Toussaint heuristics *vs* initial number of points

> *From left to right: rectangle-bound, uniformly distributed in a circle and centrally distributed in a circle. Red line is when the quadrilateral is used, blue line when it is the hexagon and black line when it is an octogon.*

From this figure it can be seen that using either a quadrilateral or an hexagon does not fundamentally change the percentage of reduction for any kind of distribution. Roughly the heuristics eliminates one half of the points. In fact it is quite frequent that extreme points meet at least two criteria, i.e. for a quadrilateral a point being at the same time the lowest *and* the right most point, or conversely the highest *and* left-most, or any other combination. Using a quadrilateral for instance frequently leads not to a quadrilateral but more often to a triangle and sometimes even to a simple line, leaving most if not all of the points outside.

However a special case appears when using an octogon on a random rectangle-bound dataset. In that case the percentage drops to approximatively the square root of the initial number of points (*exact N exponents are 0.98 for quadrilateral and hexagon and 0.46 for an octogon*).

As established by Souviron[16] rectangle-bound is the best model of random points for convex hull computation tests if it aims at reproducing the natural data distributions. Consequently from here on in this paper only the rectangle-bound model of random data will be used and "*random data*" will mean "*random rectangle-bound data*".

In order to confirm this result the same tests were run on real datatasets. They come from a variety of origins and cover a wide range in the number of points and distributions. They are formed from lightning data[1], subsets of public geo-political information files[2], medical images[3], subsets of some botanical data[4], two geographical maps[5] and computer-generated examples of clusters used for research purposes[6]. As a whole they form 790 datasets containing from 4 up to more than 760,000 points. Figure 3 displays the comparison between the Akl-Toussaint heuristics applied to random rectangle-bound datasets and real data, using for the sake of clarity only the usual quadrilateral and an octogon.

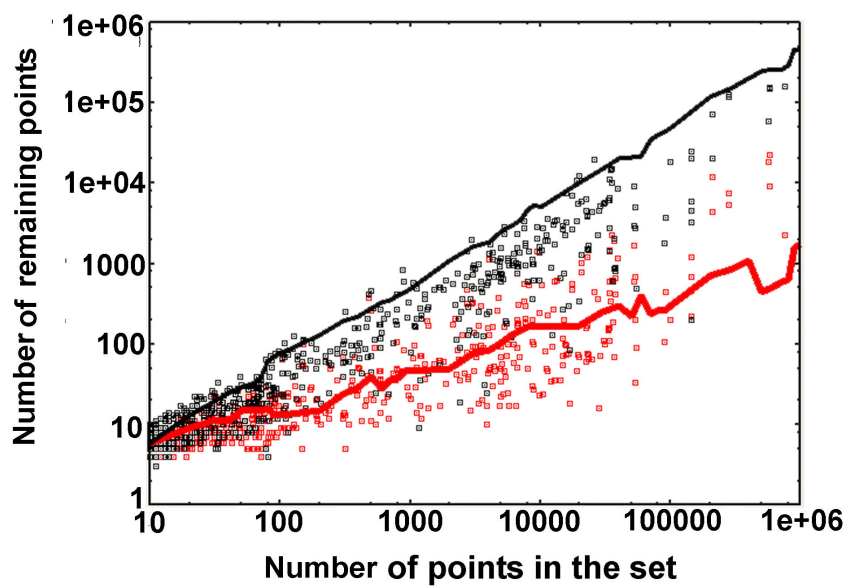

**Figure 3.** Number of points left after Akl-Toussaint heuristics *vs* initial number of points

*Lines are for the random datasets while symbols are for real datasets.*
*Black is when the quadrilateral is used and red when it is an octogon.*

Although it appears that using a quadrilateral is slightly more performant on the real datasets than on random data, the number of remaining points is still around 10% of the number of input points while these results confirm that with an octogon it is of the order of magnitude of the square root of the initial number of points.

Thus two essential conclusions to this chapter can be drawn:

- When using the Akl-Toussaint heuristics the number of initial points should be considered a parametre as these points can meet several citeria at the same time.
- For real efficiency an octogon should be used *in lieu* of a quadrilateral or an hexagon.

---

[1] Lightning strike locations obtained in two days in the summer of 1998 through the CLDN (*Canadian Lightning Detection Network*), courtesy of Environment Canada. Selected within time bins (*from 10 minutes up to 2 hours*) and resolution bins (*from 2.5 up to 350 km minimum distance between locations*), they form a sample of 629 datasets, ranging from 4 to more than 93,000 points.
[2] RGC dataset (*France's cities geographic directory*) of IGN (*french National Geographic Institute*). 31 files were obtained by selecting several population ranges as well as several city's area ranges.
[3] 10 grainy images of most of the categories of the 2D Hela databank of the US National Institute of Aging were thresholded to various high levels as to obtain 88 files of irregular and separated points.
[4] Cover dataset from the UCI Machine Learning Datasets Repository. 16 files were obtained by selecting the different cover types (*extreme density*).
[5] High resolution (*down to 10-metres accuracy in some areas*) hydrological network and coastal map of North America courtesy of Environment Canada.
[6] 24 clustering datasets of the Speech and Image Processing Unit at the University of Eastern Finland

# 3. Akl-Toussaint-based convex hull computation

The implementation of a convex hull computation using the Akl-Toussaint heuristics is in two steps:

- Apply the Akl-Toussaint heuristics to the initial dataset to reduce the number of points
- Compute the convex hull using any kind of method on this reduced dataset.

### 3.1 Basic drawback in the Akl-Toussaint heuristics

The basic flaw in the Akl-Toussaint heuristics lies in the fact that in order to keep or eliminate a point, tests are made on *all* the box's vertices, whichever point-in-polygon algorithm is used. If a point is to be kept these tests are then thrown out as the only result is "*the point is outside the box*". In the original paper the authors mentioned the dispatching of the points into four regions only *after* the interior points were discarded. However in reality these tests can provide additional information which can be of great help for the building of the convex hull, as it will improve the overall efficiency of the algorithm.

### 3.2 Avoiding this drawback

It all starts with a basic observation: through its geometric philosophy the Akl-Toussaint heuristics provides for a way of optimising the search for new convex vertices. *By construction* and if the initial points are sorted in an anti-clockwise manner possible candidates lie on the *right* side of one of the original box's segments, and more precisely on the right-side of *only one* segment (*see Figure 4*): this heuristics divides the space into regions relating to *only* one side of the original box.

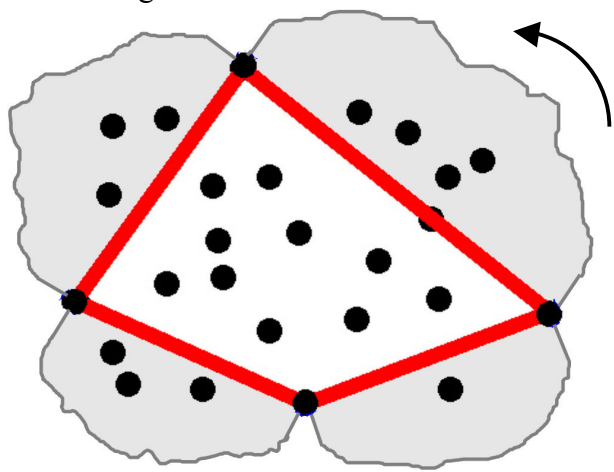

**Figure 4**: Space partitioning of points by the initial box

*Red lines are for the initial box. Gray areas define the right side of each segment. Here is the case with 4 points, but it is the same if another combination is taken (6 or 8) The arrow defines the anti-clockwise order.*

As a point is inside the original box if it is left of *all* box's segments, then if the point is right of *one* segment it is potentially a new convex vertex located in between this segment's extremities.

In doing so one uses the opportunity given by the throw-away checking to find at the same time within which vertices of the original box this point might be included.

Now in order to keep the benefit of this new information before processing the next point an incremental approach is the obvious choice, as no additional storage is required nor is further processing later on, and the convex hull is updated as soon as a new outside point is found

However one problem remains: the original segment to which this point is related has been identified. But if the convex hull has already been updated some vertices might have been inserted in between the segment's extremities, and thus the hull should be explored to.check whether the point is actually inside the updated hull and if not, its exact insertion location. A simple solution to that problem lies in storing the indexes of the original box points and updating them for each new inserted vertex.  It thus limits the exploration to the *range* of indexes for this segment. And now, as Figure 5 shows, the same kind of checking can be made for this range: if the point is on the right side of a convex segment of the updated hull it has to be inserted within the boundaries this segment's limits.

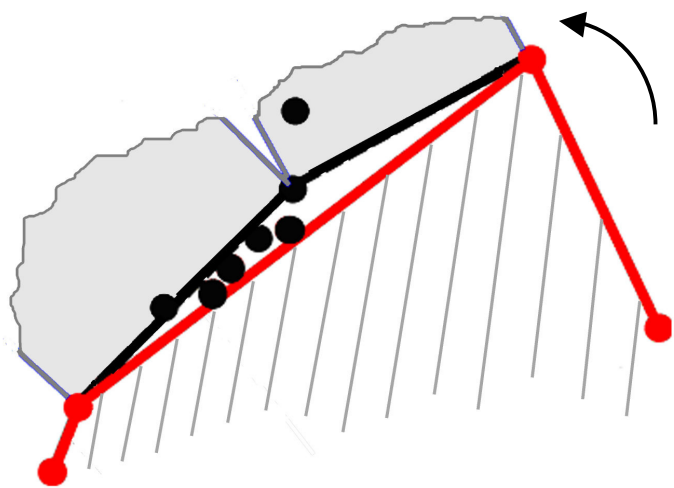

**Figure 5**: Relative positioning of possible candidates versus subsequent vertices

*Red lines are for the initial box. Black lines represent the temporary hull once a point has been inserted. Gray areas define the limits of the 2 sub-segments once one vertex is included between the 2 box's vertices The shaded area is the interior of the original box. The arrow defines the anti-clockwise order.*

Although in Figure 5 it can be seen that the point is on the right-hand side of more than one sub-segment of this part of the convex hull, this will be dealt with later on.

In conclusion, using this approach allows first to find which side of the initial box is closest to the point at the same time that it determines whether the point is inside or outside, and secondly which actual edge of the updated hull is closest by exploring only vertices situated in between the original vertices already found. It thus optimises the search and checks.

Moreover as in all incremental methods forward and backward checks have to be performed but this checking is also limited to the vertices situated in between the already found limits (*thus solving the above-mentioned ambiguity*).

The only drawback is that a small (*the size of the defined number of characteristic points*) array has to be used to store the positions (*or the addresses*) of the intial vertices.

So the algorithm goes as follows:

**1.** The initial box is computed according to the authors's guidelines for octogons (*i.e. points corresponding to extreme x and y, as well as to extreme sum and difference of coordinates*) and the indexes of its vertices in the global array of points are stored, in counter-clockwise order, as well as the vertices themselves.

**2.** The algorithm then loops through all points in the data set and checks whether each point is inside or outside this box. If a point is found on the right of a segment the related box's segment is found.

**3.** For such a point, the algorithm then checks through the sub-segments defined by the range covered by the box's segment extremities and finds whether it lies on the right side of one. If the point lies on the left-hand side of all the sub-segments, the point is *outside* the original box but *inside* the temporary convex hull and the algorithm then iterates to the next point. Otherwise the point is outside the temporary convex hull hence it is a new temporary convex vertex. Forward and backward checking are then made and vertices eventually deleted, then the point's index is inserted in the temporary hull (*and the box indexes are eventually updated as well if an array of indexes is the chosen data structure*).

The algorithm in pseudo-code is thus straightforward and shown below, *nbox* being the chosen number of box's corners (*here 8*):

```
Finds the initial box's corners in anti-clockwise order

Loop using i from 1 to N

    Loop using j from 0 to nbox

        If Pi is right of the j^th box's segment

            Loop for all segments within the limits for this segment

                If Pi is right of one segment

                    Checks backwards for concave angles until low limit
                    Checks forwards for concave angles until high limit
                    Inserts Pi
                    Exits

                EndIf

            EndLoop

            Exits

        EndIf

    EndLoop

Endloop
```

It is worth noticing that as the chosen number of characteristic points *might not be* the actual real number a variable *nbox* is used in the pseudo-code.

### 3.3 Space Complexity

This algorithm is space-efficient as it needs only the final number of output points as internal storage. If only the indexes of the convex vertices are needed it is in O(*1*) space-wise or O(*h*) if the vertices are to be output as an array of points.

### 3.4 Time Complexity analysis

Let *N* be the number of points, *p* the number of the real initial box's vertices, *p0* the expected (*set*) number of characteristic points, *N'* the number of points outside the initial polygon, and *h* the final number of convex vertices.

- Finding the initial box's vertices requires $N\,p0$ operations.
- Finding whether the points belong to the initial polygon requires $N\,p$ operations at most (*one can stop the exploration as soon as the point is right of one segment*).
- For a potential candidate, as there will be $h/p$ vertices lying between two consecutive box's vertices in average, finding the related segment for the whole set requires at most (*like above one can stop as soon as the point is right of one segment*) $N'\,h/p$ operations in average.
- Checking backwards and forwards requires also $(h-p)\,h/p$ operations at most: given the average number of segments lying in between two consecutive box's vertices, one has potentially to check for half of this range, i.e. $h/2p$ vertices in average in each direction. Experimentally though it appears that usually there is an average of only one point explored through backward or forward checking.

Thus the total number of operations is: $N\,(p+p0) + N'\,h/p + (h-p)\,h/p$     **(1)**.

The dominant factor in equation (1) is either $2\,N\,p$ or $N'\,h/p$ depending on the distribution of points (*supposing that p equals p0*), as the last factor is smaller or equal to the second one.

In the worst-case for which all points are part of the convex hull, *N'* equals $(N-p)$, *h* equals *N*, and *p* equals *p0*, the total number of operations is: $2\,Np + 2\,(N-p)\,N/p$, which is dominated by $2\,N^2/p$ or $O(N^2)$ (*which however will be p/2 times faster than Jarvis's march worst-case*).

In the general average case and if an octogon is intially chosen the second term will dominate over the third (*N' is greater than h*), and it will dominate over the first one if *h* is superior to 128. However numerical values taken from Souviron´s previously cited paper suggest that this value will only be reached for sets above *at least* $1.2\ 10^9$ points for real datasets. But as for octogons it was established in Chapter 2 that $N' \sim 2\sqrt{N}$ this threshold will only be reached after $2.4\ 10^{18}$. Below this number the computation will then entirely depends upon *p*, leading to a linear behaviour with a constant factor growing from 16 to 32 at a rate of $6.67\ 10^{-18}$.

### 3.5 Improvements

Although equation (1) is correct in theory it depends on the chosen data stucture: if points or indexes are represented as an array the insertion/deletion costs should be added. In average it will add an $(h-p)\,h/2$ factor, as for each new point added one must shift the remaining part of the array, half of which will be concerned in average. This will then lead to equation (2):

$$N\,(p+p0) + N'\,h/p + (h-p)\,h/p + (h-p)\,h/2 \quad \textbf{(2)}.$$

On the other hand, using a doubly chained list does not have this inconvenient but uses three times more memory.

An obvious global improvement would be to use a dichotomy to search for the right sub-segment within the limits of the found box's segment. This will reduce the second term of equation (1) to $N'$ log($h/p$). Unfortunately dichotomy is impossible on a doubly chained list. For arrays structures however if will improve the efficiency especially on the worst-case scenario.

Another improvement for the array structure could be to dispatch the temporary vertices to separated arrays, one for each side of the initial box. This will improve on the last factor in equation (2) (*the memory shifts*) by reducing the $h$ factor to $h/p$ in average.

Figure 6 below shows the relative gain using each or a combination of these improvements for both the average and worst-case scenarii.

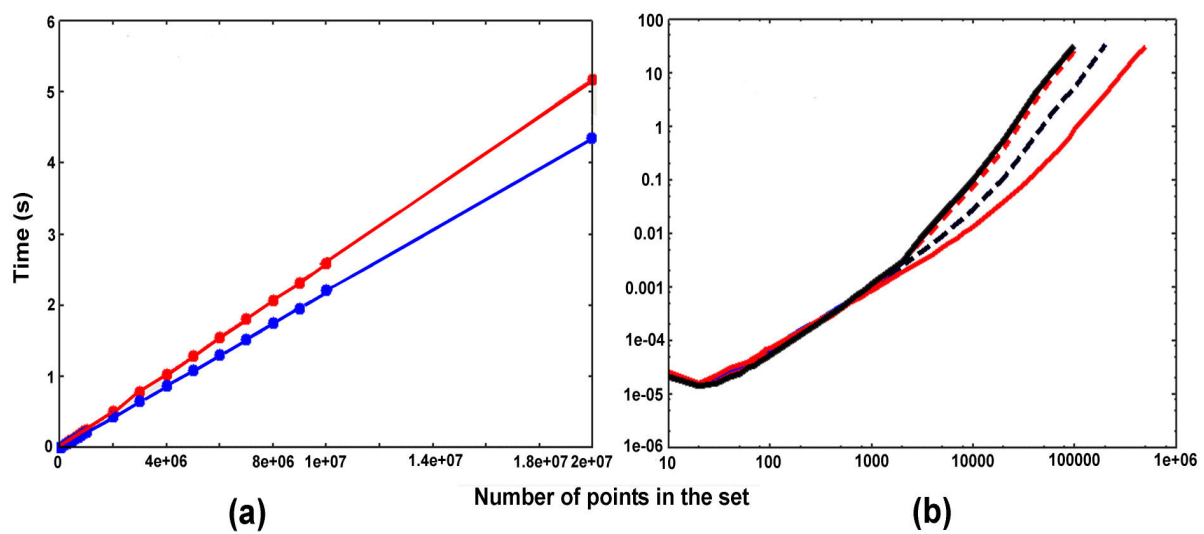

**(a)**

**(b)**

**Figure 6**: Comparison of the different improvements techniques for the array structure

(a) *Average case: red line is the basic case, blue line repesents the dispatch amongst different buffers. Dichotomy effects are un-noticable until $9 \cdot 10^6$ points.*

(b) *Worst-case: black line is the basic implementation, dashed red line is the dispatch to the different buffers, dashed black line is using the dichotomy and solid red line represents using both the dichotomy and the dispatch to the different buffers.*

For the average case the main improvement lies in the dispatch to the different buffers as it minimses the memory movements, as can be seen in Figure 6a, while for the worst-case scenario the combination of the two improvements leads to a better result as expected.

As a conclusion to this part it must be stressed that there are no possible improvement for the algorithm if chained lists are used: although they will produce space-efficient and optimal speed in the average case (*as no memory shift will be involved*) they will still run in O($N^2$) in the worst-case scenario as no dichotomy can be used to improve the search amongst sub-segments of an initial box's side. Hereafter "*optimised version*" will refer to the array-structure version using both dichotomy and dispatch to the different buffers. To summarize either using arrays or chained lists will produce an O($N^2$) worst-case time complexity as each of these data structure has its own drawback but will be linear in the average case up to 1.2 billion points as was expected from Devroye & Toussaint[8].

### 3.6 Comparison with other convex hull algorithms

In order to establish the efficiency of this algorithm and its variations four well-known convex hull algorithms were tested: a gift-wrapping algorithm, a Graham Scan, a quickhull, and an algorithm based on Andrew[3]'s monotone chain algorithm by Clarkson[6]. Only the basic implementation and the most optimised one are shown here as not to clutter the plots.

Although running time comparisons here are only relative technical information is as follows: code was written in C, running under Linux CentOS4.6, using a 32-bits monoprocessor AMD[R] Sempron 2.8 GHz and 256 MB DDR (*Acer Aspire 1362WLMi*). All coordinates were floating point numbers (*double precision*)

First in order to prove the linearity and establish the relative performance of even the basic implementation with arrays a linear comparative plot is in Figure 7 then parallel worst-case comparisons are shown in Figure 8 using log-log scales.

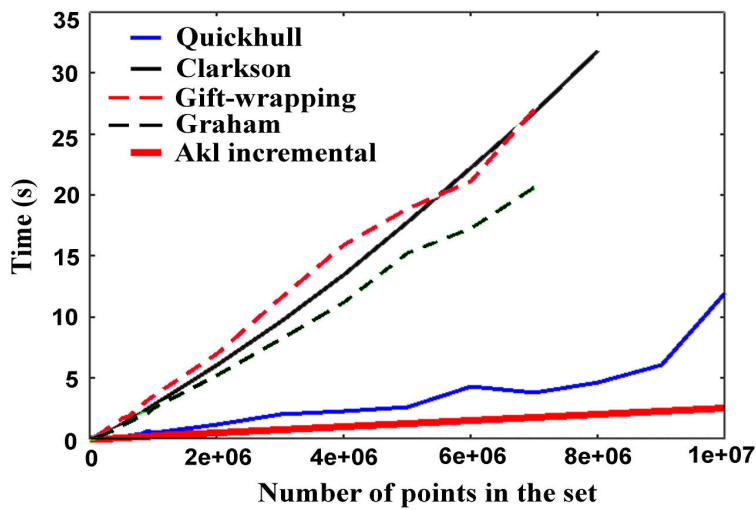

**Figure 7**: Performance comparison on average case using the basic implementation

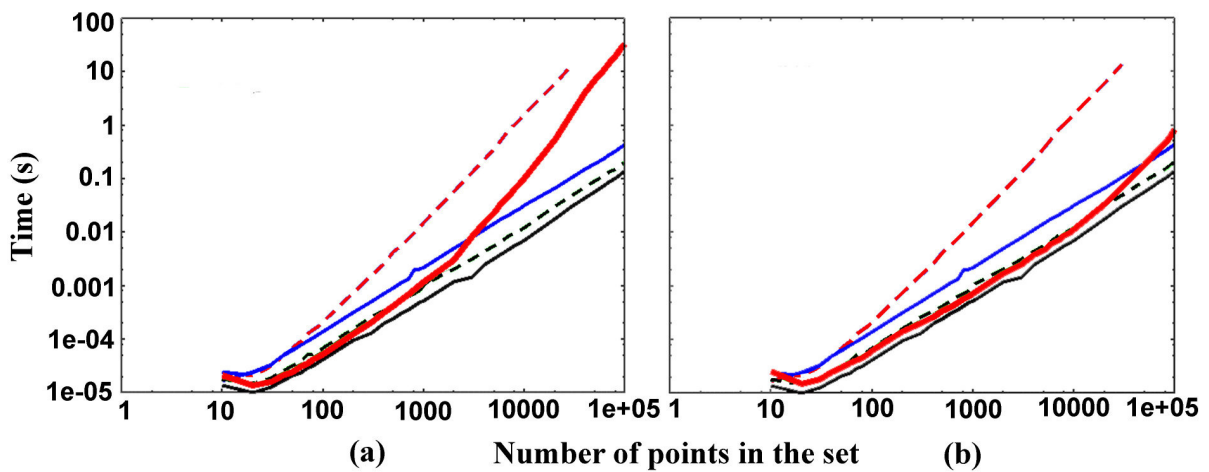

**Figure 8**: Performance comparison on worst-case scenario

*(a) basic version (b) optimised version.*

*Curve identificators are the same than above.*

Finally comparisons were made between the optimised version and combinations of the Akl-Toussaint heuristics with one of the algorithms as it is usually done. To put things on a comparative basis both the usual quadrilateral and the octogon were chosen during the Akl-Toussaint elimination process and were also chosen for the incremental versions. Results are shown in Figure 9.

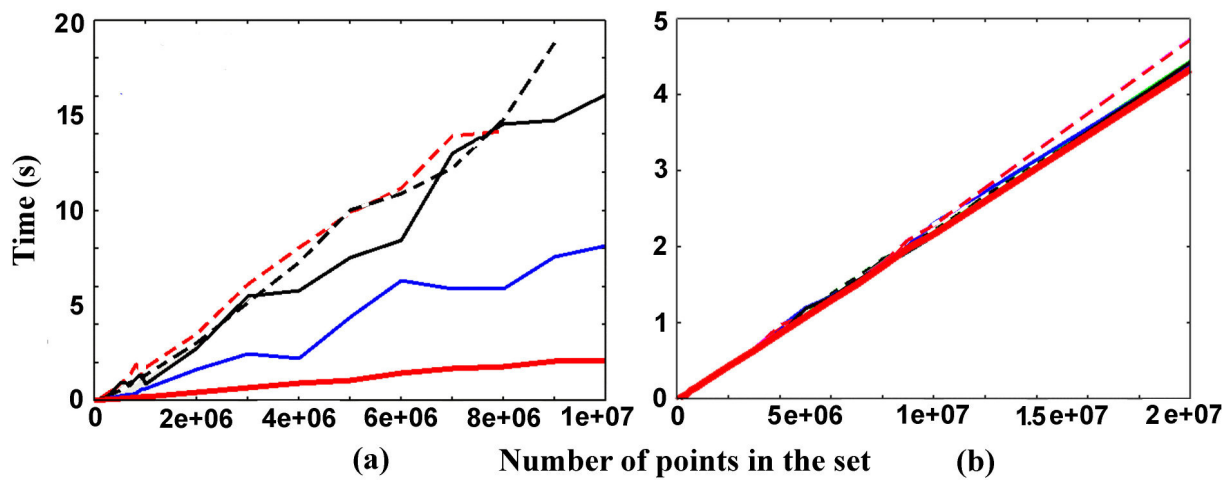

**Figure 9**: Performance comparison on average-case scenario

*(a) Using a quadrilateral (b) Using an octogon*

*Curve identificators are the same than in Figure 7.*

It is worthwhile noticing from Figure 9a that a quadrilateral-based optimised incremental implementation runs in the same amount of time than the octogon-based implementation in Figure 7 as the gains on the $p$ factor are compensated by the loss on $N'$ and consequently more operations

To conclude this chapter it can be said that in the average case the incremental versions performs better than the tested ones even if they are combined with an equivalent Akl-Toussaint heuristics elimination process. Also, depending on the average size of expected worst-case situtations even the most basic algorithm might be better than a straighforward use of other methods (*i.e. with no combined heuristics-based elimination*). Moreover these methods are space-efficient.

## 4.  Note on the incremental method

The incremental convex hull computation is a basic well-known method, the first one to come to mind and its complexity analysis was upper-bounded by Kallay.

The algorithm starts by creating an initial hull with the first 3 non-aligned points in the data set. Then, exploring the array, it checks whether the next point is inside this box (*this is equivalent to using Akl-Toussaint heuristics on a changing box*). Finally checks are made forwards and backwards to verify whether the new point will produce concave angles on either sides, then this point is added to the hull, and so on.

The advantage of this method is to directly build the convex hull without using any additional computations whether it be sorting or elimination although most authors mention pre-sorting before using it. It is also space-efficient as no additional storage other than the vertices themselves is required (w*hich can even be avoided if in-place computation is allowed*) and well suited for on-line computations.

However *all* intermediate convex vertices *might not* be part of the final convex hull. So the backward and forward checks in this approach have to allow for a full checking on all already computed convex vertices.

As the hull is evolving with time, it is difficult to compute overall efficiency in details. Using the same notations as above and denoting *h'* the temporary number of vertices at one stage, then for each point in the array:

- Finding whether it belongs to the temporary polygon requires at most *h'* operations
- Checking backwards and forwards also requires also *h'* operations at most in average.

Thus the total number of operations for one point is $2h'$ in average. Now *h'* can be temporarily superior to the final number *h*. Experimentally though it appears that *h'* never exceeds $h+2$, and only a few times at that. Also, like above-mentioned, there is usually an average of only one point explored through backward or forward checking. However one might say that the average value for the processing of the whole set is $2hN$ i.e. in O*(Nh)*. In the worst-case where *h* equals *N* the total number of operations is therefore dominated by $2N^2$ or O($N^2$).

Even in a straightforward array-based implementation it will for average cases be almost twice as fast than applying an Akl-Toussaint heuristics for it will save one loop on *N*, although memory movements will be in O(*h*) rather than in O(*h/p*) and search in O(*log h*) rather than in O(*log h/p*), as Figure 10 shows (*in view of Figures 7 and 9b the octogon-based optimised incremental version is used as a reference*). Using above-mentioned numerical values it could be said to be in O($N^{1.09}$) in average with a very low constant factor.

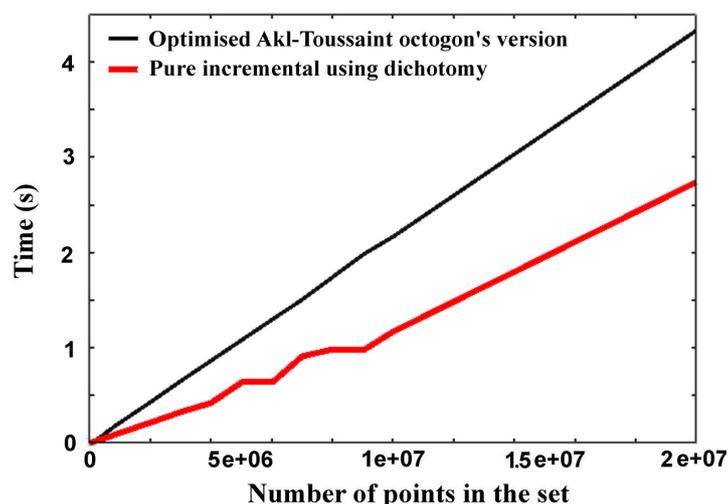

**Figure 10**: Performance comparison on average-case scenario for the purely incremental method

Finally several publications focused on a probabilistic approach of "*randomized incremental hull computation*" like Clarkson et al[7], artificially randomizing the datasets, whereas in the general case points *are* in random locations. However, setting aside the theoretical aspect, as software programs seldom run on worst-case data and Nature does not either frequently provide worst-case data, from a physicist's point of view one has to wonder at the widespread disregard and even disdain towards this basic method in spite of its space and computational efficiency and its simple coding.

# 5. Optimal algorithm

## 5.1 Algorithm's principle

As previously mentioned the weak point of all the above-mentioned incremental methods lies in their worst-case behaviour.

However Chan's idea was to guess a value for $h$ and divide the original set of points into $h$ buckets in order to reach optimal complexity. Using Akl-Toussaint heuristics naturally divides the set of points into $p$ buckets while eliminating a great number of points in the average case. Using Chan's idea one can then compute the convex hull of each of these buckets using a O($N$ log$N$) algorithm. Thus using the same notations than above the total number of operations required to process the whole set of points is given by equation (3):

$$2\,Np + \mathrm{p}\,(N'/p \log N'/p) \quad \textbf{(3)}$$

The second term will dominate the first when log $N'/p$ is greater than $2p$. If an octogon is chosen this will happen when $N'$ equals $1.6\ 10^{17}$ or $4\ 10^{8}$ if it is a quadrilateral. These are limits for the worst-case scenarii while given the above-mentioned numerical values of $N'$ as a function of $N$ however in the average scenario the thresholds will thus be $3.2\ 10^{34}$ for an octogon or $1.6\ 10^{17}$ for a quadrilateral Under this number the computation will only be dependent upon $p$ and thus be linear, with a constant factor growing from 16 to 32 (*or 8 to 16*) at a growing rate of $2\ 10^{-33}$ (*or 5 $10^{15}$*). Although these thresholds are not *mathematically* infinite values they are high enough to conclude that the algorithm is O($N$) in practice.

## 5.2 Improvements

The number of points in each bucket can be reduced by using a technique similar to the one of the quickhull algorithm, but in a dynamic way: for a new point outside the initial box one might check whether it is inside the triangle formed by the initial box's segment and the point at the maximum distance from this side (*this point is dynamically modified as soon as a new outside point is found*).

The size of the set on which the O($N$ log$N$) algorithm runs can further be reduced by using a left/right division from the middle of the initial box's segment like in Kirkpatrick & Seidel, putting for instance the right part at the beginning of the bucket as it will be processed first.

Finally memory operations could be costly. A truly space-efficient version would re-allocate the buckets as needed. In the advent of a worst-case scenario however this will be time-consuming. On the other hand allocating from the start $N/p$ would be much too large for average cases. A middle road lies in pre-allocating each bucket to the expected size of the average case, i.e. to $2\sqrt{N/p}$ .

Combined together these improvements produce the best balance of results on both the average case and the worst case as they reduce the growth rate of the constant factor. Although the second improvement is fast to compute the computations involved in the first one might hinder the performances for worst-case scenarii but improve the average case. Thus depending on situations one might choose *not* to implement the first of these improvements.

### 5.3 Pseudocode

The algorithm in pseudo-code is thus straightforward and shown below, *nbox* being the chosen number of box's corners (*here 8*):

```
Finds the initial box's corners in anti-clockwise order

Loop using i from 1 to N

    Loop using j from 0 to nbox

        If Pi is right of the jth box's segment

            Eventually applies one of the improvements to further restrict
            Stores the point in the jth buffer
            Exits

        EndIf

    EndLoop

Endloop

Loop using j from 0 to nbox
    Computes the convex hull of the jth buffer using an O(NlogN) algorthm
Endloop

Loop using j from 0 to nbox
    Stores initial point j
    Stores the convex vertices from the jth buffer
Endloop
```

### 5.4 Comparisons with other algorithms

Finally comparisons were made between this algorithm and other algorithms combined with an octogon-based Akl-Toussaint elimination process like what was done in Chapter 3. Results are shown in Figure 11.

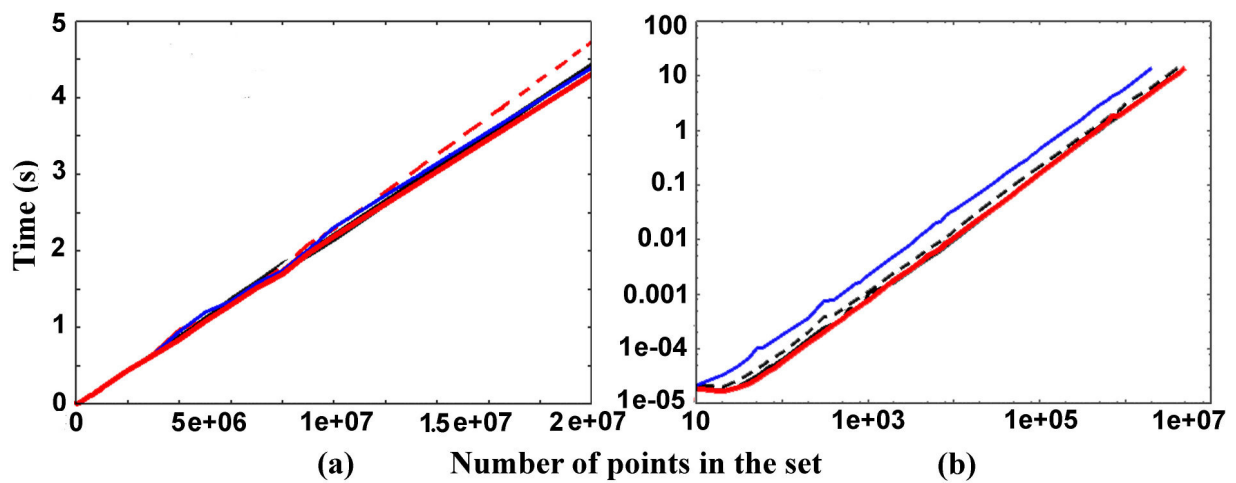

(a)      Number of points in the set      (b)

**Figure 11**: Performance comparison of the optimal algorithm

*(a) Average case (linear scales) (b) Worst case (log-log scales)*

*Curve identificators are the same than in Figure 7.*

## 6. Conclusion

First a study of the Akl-Toussaint heuristics's elimination process as a function of the number of initial characteristic points established that using an octogon is much more efficient than the usual quadrilateral as it leaves only O($\sqrt{N}$) points. Then simple-to-implement incremental space-efficient convex hull algorithms were presented. They are modified versions of the Akl-Toussaint heuristics allowing for a full convex hull computation and not only as a means to reduce the number of points. Their main advantage lies in their simplicity and efficiency both in memory and speed as they run in O($N$) in the average case although their worst-case is in O($N^2$). A remark was then made on why for average cases a simple incremental should be preferred. Finally an optimal algorithm was derived with an O($N$) time complexity whether in the average or worst case. It mixes the working-from-inside approach with more well-known algorithms and is space-saving if not space-efficient.